\documentclass[conference]{IEEEtran}
\IEEEoverridecommandlockouts

\usepackage{cite}
\usepackage{amsmath,amssymb,amsfonts}
\usepackage{algorithm}
\usepackage[noend]{algpseudocode}
\usepackage{graphicx}
\usepackage{textcomp}
\usepackage{xcolor}
\usepackage{multirow}
\usepackage{hyperref}
\usepackage[justification=centering]{caption}
\usepackage{subcaption}
\usepackage{pifont}
\usepackage{listings}
\usepackage{bm}
\usepackage{makecell}
\usepackage{booktabs}
\usepackage{float}
\usepackage{tablefootnote}

\usepackage{soul}

\DeclareMathOperator*{\argmax}{argmax}   

\newcommand{\cmark}{\ding{51}}%
\newcommand{\xmark}{\ding{55}}%

\begin{document}

\interfootnotelinepenalty=10000

\title{SAMO: Optimised Mapping of Convolutional Neural Networks to Streaming Architectures}

\author{
    \IEEEauthorblockN{Alexander Montgomerie-Corcoran\IEEEauthorrefmark{1}, Zhewen Yu\IEEEauthorrefmark{1} and Christos-Savvas Bouganis}
    \IEEEcompsocitemizethanks{\IEEEcompsocthanksitem\IEEEauthorrefmark{1}\textit{equal contribution}}
    \IEEEauthorblockA{Dept. of Electrical \& Electronic Engineering \\ 
    Imperial College London, UK \\
    \{alexander.montgomerie-corcoran15, zhewen.yu18, 
    }
}

\maketitle

\begin{abstract}
Significant effort has been placed on the development of toolflows that map Convolutional Neural Network (CNN) models to Field Programmable Gate Arrays (FPGAs) with the aim of automating the production of high performing designs for a diverse set of applications. However, within these toolflows, the problem of finding an optimal mapping is often overlooked, with the expectation that the end user will tune their generated hardware for their desired platform. This is particularly prominent within Streaming Architecture toolflows, where there is a large design space to explore \cite{venieris_fpgaconvnet_2018,duarte_fast_2018,blott2018finn}.
In this work, we establish the framework SAMO: a Streaming Architecture Mapping Optimiser. SAMO exploits the structure of CNN models and the common features that exist in Streaming Architectures, and casts the mapping optimisation problem under a unified methodology. Furthermore, SAMO explicitly explores the reconfigurability property of FPGAs, allowing the methodology to overcome mapping limitations imposed by certain toolflows under resource-constrained scenarios, as well as improve on the achievable throughput.
Three optimisation methods - Brute-Force, Simulated Annealing and Rule-Based - have been developed in order to generate valid, high performance designs for a range of target platforms and CNN models. Results show that SAMO-optimised designs can achieve 4-20x better performance compared to existing hand-tuned designs. The SAMO framework is open-source:
\url{https://github.com/AlexMontgomerie/samo}.
\end{abstract}

\section{Introduction}
The success of deep learning models, primarily in the form of CNNs, has fuelled research into custom hardware accelerators tuned for specific models. A popular type of platform for these accelerators are FPGAs, as their versatility and range of sizes suit a variety of applications. Many toolflows have been developed that reduce the time to develop such FPGA-based systems, and the architectures generated by these toolflows generally come under two categories: Systolic Array and Streaming (dataflow) architectures \cite{venieris2018toolflows}.

In a Systolic Array architecture design, the convolution layers are mapped to a matrix multiplication engine, resulting in the processing elements of the architecture being time-shared across layers. As such, Systolic Array architectures have the ability to execute nearly any CNN model. Due to this flexibility, Systolic Array architectures are often served as compute kernels in general-purpose neural network accelerators that are not tailored to specific CNN models \cite{jouppi2017datacenter, vink2020caffe}.

Streaming Architectures, on the contrary, tailor their hardware towards the computation and memory workload of a specific CNN model, with the promise of greater performance over the Systolic Array architectures due to their customisation for the specific load. Instead of time-sharing processing elements between layers,
each layer is mapped to a custom computation kernel tailored to the characteristics of the layer, where the overall CNN computation is achieved by connecting these kernels together in a chain. As such, Streaming Architectures have high throughput and energy efficiency \cite{umuroglu2017finn} compared to equivalent Systolic Array architectures.

However, the design process for a Streaming Architecture accelerator is often more time consuming and tedious, as the customisability brings a large design space to explore. Unlike Systolic Array architectures where the exploration can be carried out by brute force enumeration using analytical resource models \cite{wei2017automated}, or roofline models \cite{zhang_optimizing_2015}, Streaming Architectures require a more informed method for efficiently exploring the design space and identifying high performance designs.

Early work on Streaming Architecture accelerators explored the design space manually, resulting in many cases in sub-optimal designs \cite{dicecco2016caffeinated, duarte_fast_2018}. Several works have proposed algorithms that allow automatic design space exploration \cite{li2016high, zhang2018dnnbuilder, venieris_fpgaconvnet_2016}, by following certain guidelines and rules that contribute towards a high performance design. However, those guidelines and rules are closely coupled to the specific accelerator framework and do not generalise to other toolflows.


This paper introduces the SAMO framework, which addresses the optimisation stage of mapping CNN models to Streaming Architectures in a unified manner. Key innovations of the tool are the introduced abstract representation for capturing the characteristics of the accelerator's building blocks - performance and resource requirements - and the Hardware Description Graph (HD-Graph), a data structure that allows the optimisation of the CNN mapping to a target FPGA device. Furthermore, the proposed framework makes explicit use of the reconfiguration feature of FPGA devices, and introduces a partitioning methodology that allows CNN mapping toolflows to achieve high throughput designs, as well as to produce valid designs even in a resource constraint setting. 

\section{Background}
A number of Streaming Architecture toolflows have been developed over the recent years targeting FPGA devices \cite{venieris2018toolflows}. Among those, fpgaConvNet \cite{venieris_fpgaconvnet_2018}, FINN \cite{blott2018finn} and HLS4ML \cite{duarte_fast_2018} are good representatives of that set due to their popularity within the research community, as well as their wide variation in the accelerator design space that they define. These three toolflows are referred to as backends throughout the rest of this paper.




FINN was originally customised for low-precision networks, specifically Binarised Neural Networks \cite{umuroglu2017finn}, but has now been extended to support CNN models with higher precision \cite{faraone2018customizing}. The core computation unit in FINN is the Matrix-Vector Threshold Unit (MVTU) which contains multiple Processing Elements (PE) and SIMD lanes for parallel Matrix-Vector operations. Regarding the optimisation of the mapping of a CNN to an FPGA, the number of PE and SIMD for each MVTU are refined by iteratively allocating extra resources to the slowest MVTU in the whole accelerator \cite{blott2018finn}. However, the authors of FINN also state this algorithm to be sub-optimal and can often be outperformed by hand-tuning\footnote{\url{https://github.com/Xilinx/finn/blob/main/src/finn/transformation/fpgadataflow/set_folding.py}}. FINN has recently been extended into multi-FPGA execution where Integer Linear Programming is used to balance the workload between multiple FPGAs \cite{alonso2021elastic}. However the design inside each FPGA is still manually tuned.

HLS4ML sacrifices the configurability of the accelerator to obtain simple and low-latency designs \cite{aarrestad2021fast}, as the tool was originally developed to accelerate shallow and simple networks for particle physics experiment \cite{duarte_fast_2018}. Therefore, HLS4ML has a relatively small design space compared with other Streaming Architecture toolflows. Specifically, HLS4ML supports two hardware generation objectives: \textit{resource} and \textit{latency}. For the \textit{latency} objective, the design architecture is completely unrolled, and the HLS compiler is given the task of optimising the latency for a given initiation interval target. In the case of the \textit{resource} objective, the hardware is only partially unrolled, and the performance of the design is manually tuned through the \textit{reuse-factor} parameter, which defines how often resources are re-used. 

fpgaConvNet \cite{venieris_fpgaconvnet_2018} focuses on the configurability of the accelerator in order to achieve high performance for a range of models and platforms. This toolflow explores several degrees of parallelism in the network including \textit{coarse-grained folding} and \textit{fine-grained folding}. The toolflow further supports bitstream reconfiguration and partial reloading of weights in order to overcome limitations imposed due to resource constraints. Therefore, fpgaConvNet has the largest design space amongst the three considered toolflows. In terms of the optimisation algorithm, fpgaConvNet proposed a Simulated Annealing optimiser \cite{venieris_fpgaconvnet_2016} to explore it's design space, however this has been tailored to this specific toolflow. 

To summarise, the above toolflows provide the necessary tools to build a highly tailored streaming architecture for the execution of a CNN on an FPGA device. The parameters of the hardware building blocks define a large design space to be explored, and expose a trade-off between performance and resources. As the CNN models and FPGA devices become larger, the design space that needs to be explored for the identification of pareto-optimal designs increases, necessitating the use of automated methods for exploration. SAMO aims to provide a unified way of exploring the space across different streaming architecture toolflows, by capitalising on the characteristics of the CNN models and the streaming architectures. 



\section{Generalised Optimisation Problem}
\label{Sec:Optimisation_Problem}

This section outlines the generalised optimisation problem of tailoring a parameterised Streaming Architecture to a target CNN and FPGA pair. The aim is to efficiently utilise the available resources on the platform in order to optimise for either a throughput or latency objective, for a given network. For simplicity, this section only discusses sequential CNN models, even though SAMO can support residual networks as well.



\subsection{Hardware Description Graph}

In order to generalise the optimisation of the CNN mapping to a target FPGA device, SAMO proposed a data structure called the Hardware Description Graph (HD-Graph), which provides an abstraction of both the CNN model topology and the hardware building block implementation. The definition of the HD-Graph is as follows:

Assume the given CNN model is represented as a Directed Acyclic Graph with $L$ layers as $M = \{l_1, \ldots , l_L\}$, where $l_i$ is the $i^{th}$ layer within the CNN model. This graph has edges $E_M$ between the layers. For sequential networks, the edges are only between adjacent nodes, so $ E_M = \{ (l_1, l_2), \ldots , (l_{L-1}, l_L) \} $. 

SAMO uses the backend's parser to translate each layer of the CNN model to a computation node or set of computation nodes in the HD-Graph, where each computation node corresponds to a parameterised hardware building block implemented in the backend. The translation process for specific backends will be elaborated on in Section~\ref{subsec:parser}. After the translation, the HD-Graph is described as $H$, which contains $N$ computation nodes, $H = \{ n_1, \ldots , n_N \}$ where $n_i$ is the $i^{th}$ node within it. As the CNN models are sequential, so is the HD-Graph, and the edges of $H$ are such that $ E_H = \{ (n_1, n_2), \ldots , (n_{N-1}, n_N) \} $. 

\subsection{Partitioning}
SAMO introduces a partitioning methodology to allow backends to generate high throughput designs by reconfiguring and time-multiplexing hardware building blocks on the single FPGA device. The partitioning methodology is described in terms of cuts of the HD-Graph $H$, which transform it into multiple sub-graphs $\bm{P}$ where each sub-graph constitutes an FPGA configuration containing a subset of nodes from the HD-Graph. 


The positions of where the cuts take places are represented by the optimisation variable $\bm{C} = \{e_{1}, \ldots, e_{|\bm{P}|-1}\}$, and $|\bm{P}|$ denotes the number of sub-graphs. As such, the computation nodes from which the sub-graphs $\bm{P}$ are constructed are defined in Eq.~(\ref{eq:partitions}).

\begin{equation}
    \bm{P} = \begin{cases}
        \{ n_{e_i+1}, \ldots , n_{e_{i+1}} \} \; \forall \; e_i \in \bm{C} & |\bm{C}| > 1 \\
        \{ n_1, \ldots , n_{e_i} \}, \; \{ n_{e_{i+1}}, \ldots , n_N \}  & |\bm{C}| = 1 \\
        \{ n_1, \ldots , n_N \} & |\bm{C}| = 0
    \end{cases}
    \label{eq:partitions}
\end{equation}

This leads to the properties that the partitions are disjoint ($\bigcap \bm{P} = \emptyset$) and complete ($\bigcup \bm{P} = H$).




\subsection{Variables}
In the HD-graph, each computation node corresponds to a parameterised hardware building block whose implementation is backend-specific. To capture the possible parametrisation of the current landscape of Streaming Architecture frameworks, SAMO defines three associated variables for each node $n_i$:
\textit{input channel folding} ($s_i^I$), \textit{output channel folding} ($s_i^O$) and \textit{kernel folding} ($k_i$). These variables can be vectorised across all nodes ($\bm{s}^I, \bm{s}^O, \bm{k}$).

The \textit{input} and \textit{output channel folding} variables describe the degree of parallelism of the channel dimension of the feature-map entering and exiting a node respectively. 
The \textit{kernel folding} variable describes the parallelism of computations within a node. 

\subsection{Objective}
\label{subsec:Objective}
With all the optimisation variables defined, the optimisation problem is now outlined. SAMO supports two optimisation objectives, which are latency minimisation and throughput maximisation. For the objective, $V=\{\bm{C}, \bm{s}^I, \bm{s}^O, \bm{k}\}$ denotes the the HD-Graph configuration for the given optimisation variables.

As the hardware building blocks are pipelined and the CNN models are assumed to be sequential, the latency of each sub-graph is dictated by its slowest node\footnote{Pipeline depth is ignored as it has a negligible effect on latency.}, which is defined in Eq.~\eqref{eq:latency}.

\begin{equation}
\begin{aligned}
\mathcal{T}(P_i) &= \text{max}\{ t(n_j| s_j^I, s_j^O, k_j) : n_j \in P_i \} \\
\end{aligned}
\label{eq:latency}
\end{equation}

$t(n_j| s_j^I, s_j^O, k_j)$ is the performance model for node $n_j$, as defined by the backend. $\mathcal{T}(P_i)$ is the latency estimate for the partition $P_i$.
When considering the whole HD-graph, the objective for minimising latency of a design is described in Eq. (\ref{eq:optimisation_problem_latency}), where $t_{conf}$ is the reconfiguration time of the device.
\begin{equation}
\label{eq:optimisation_problem_latency}
\begin{aligned}
\mathcal{O}(V) = \left( \sum_{P_i \in \bm{P}} \mathcal{T}(P_i) \right) + |\bm{C}| \cdot t_{conf} 
\end{aligned}
\end{equation}

Conversely, the objective of maximising throughput for a given FPGA and network pair is described in Eq. (\ref{eq:optimisation_problem_throughput}), where $B$ is the batch size for the inputs to be run.
\begin{equation}
\label{eq:optimisation_problem_throughput}
\begin{aligned}
\mathcal{O}(V) = - \frac{B}{B\cdot \left( \sum_{P_i \in \bm{P}} \mathcal{T}(P_i) \right) + |\bm{C}| \cdot t_{conf} } 
\end{aligned}
\end{equation}

The above two optimisation objectives can be used to construct the following optimisation problem,
\begin{equation}
    \min_{V} \mathcal{O}(V),\;\;V=\{\bm{C}, \bm{s}^I, \bm{s}^O, \bm{k}\}
\end{equation}
where $\mathcal{O}$ represents the objective, either total latency or negated throughput.


\subsection{Constraints}
\label{subsec:constraints}
In order to generate a valid and synthesisable accelerator design, there are certain constraints imposed on the optimisation problem. In this subsection, we define the constraints that are observed across all backends, although specific backends are not necessarily constrained by all.

The first constraint defined is on resources in Eq. \eqref{eq:rsc_constraint}. This constraint says that each partition must fit within the FPGA resource constraints.

\begin{equation}
\begin{aligned}
\mathcal{R}(P_i) = \sum_{n_j \in P_i} r(n_j| s_j^I, s_j^O, k_j) \\
\max\{ \mathcal{R}(P_i) : P_i \in \bm{P} \} \leq \mathcal{R}_{platform} \\
\end{aligned}
\label{eq:rsc_constraint}
\end{equation}

$r(n_j| s_j^I, s_j^O, k_j)$ and $\mathcal{R}(P_i)$ denote the resource utilisation per node and per partition respectively. The resource types considered include DSP, BRAM, LUT, FF.

A further constraint is placed on the bandwidth it must be less than the memory bandwidth for a given partition, as described in Eq. \eqref{eq:bandwidth_constraint}.
\begin{equation}
\begin{aligned}
    \mathcal{B}(P_i) = \frac{\mathcal{D}^{I}(P_i) + \mathcal{D}^{O}(P_i)}{T(P_i)} \\
    \mathcal{B}(P_i) < \mathcal{B}_{platform} \; \forall \; P_i \in \bm{P} \\
\end{aligned}
\label{eq:bandwidth_constraint}
\end{equation}

where $\mathcal{D}^{I}(P_i)$ and $\mathcal{D}^{O}(P_i)$ are the folded dimensions of the feature-map in and out of the partition respectively, $\mathcal{B}(P_i)$ is the bandwidth model for the partition, and $\mathcal{B}_{platform}$ is the memory bandwidth of the given platform.

To avoid padding, the channel folding variables must be factors of the channel dimension of the feature-map that the node is operating on, which is referred to as the \textit{channel factor} constraint. This is described in Eq. (\ref{eq:channel_folding_constraints}), where $c^I(n_i)$ and $c^O(n_i)$ are the input and output channel dimensions of the feature-map of the $i^{th}$ node respectively.
\begin{equation}
\label{eq:channel_folding_constraints}
\begin{aligned}
c^I(n_i) \bmod s_i^I &= 0 \; \forall \; n_i \in H \\
c^O(n_i) \bmod s_i^O &= 0 \; \forall \; n_i \in H \\
\end{aligned}
\end{equation}

Furthermore, there are certain types of layers, such as Max Pooling or ReLU layers, where the output channel dimension depends on the input. Therefore, for the computation nodes corresponding to these type of layers, $H' \subset H$, they must have matching \textit{input channel folding} ($s_i^I$) and \textit{output channel folding} ($s_i^O$). This equality constraint is described in Eq. (\ref{eq:channel_folding_matching_constraints}), which is referred to as \textit{intra folding matching}.
\begin{equation}
\label{eq:channel_folding_matching_constraints}
s_i^I = s_i^O \; \forall \; n_i\in H'
\end{equation}

The constraint of matching folding factors may also exist between nodes in order to ensure that all the data lines are connected. This constraint is described in Eq. (\ref{eq:inter_matching}), and is referred to as \textit{inter folding matching}. 
\begin{equation}
\label{eq:inter_matching}
s_i^O = s_{i+1}^I \; \forall \; n_i \in H
\end{equation}


\begin{figure*}[t]
    \centering
    \includegraphics[width=0.9\textwidth]{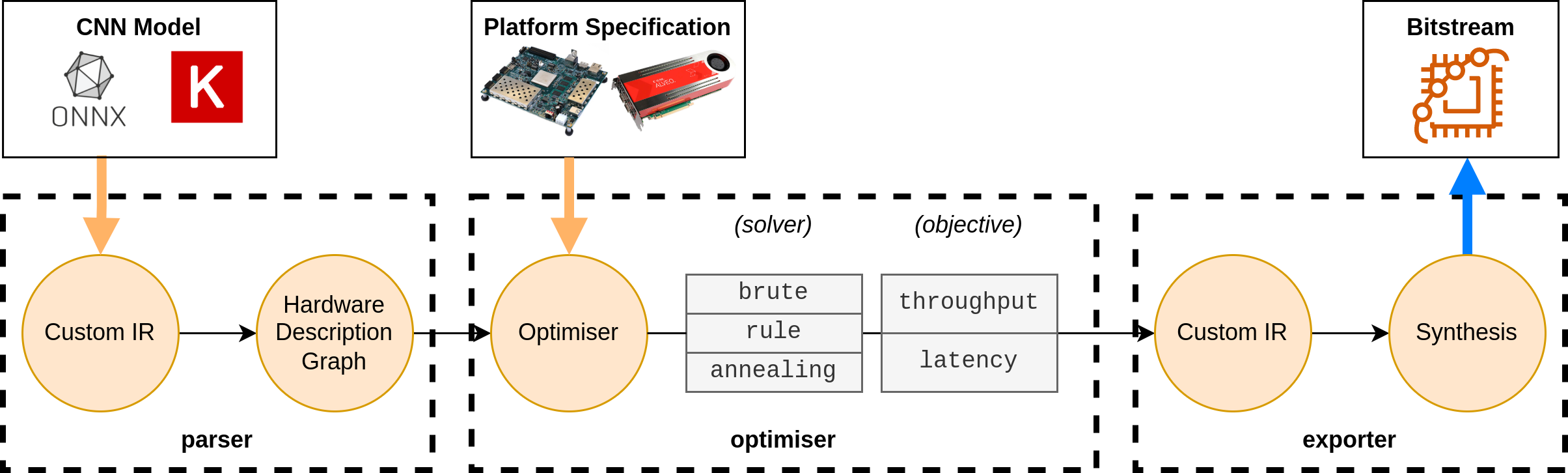}
    \caption{Overview of the proposed SAMO framework}
    \label{fig:SAMO_framework}
\end{figure*}

\section{Backend Integration \& Optimisation Schemes}
\label{sec:framework}
So far, the constrained optimisation problem has been defined, which is applicable to all the backends thanks to the abstraction that SAMO provides. In this section, we elaborate on how the proposed SAMO framework integrates the backends including fpgaConvNet, FINN and HLS4ML, and subsequently how SAMO solves the optimisation problem. An overview of our framework is given in Figure \ref{fig:SAMO_framework}, which highlights the key components of SAMO including the parser, optimiser and exporter.


\subsection{Parser: customised IR to HD-graph}
\label{subsec:parser}
The existing backends take the CNN model as an input and transform it into their own customised Intermediate Representation (customised IR), which bridges the gap between the network topology and hardware implementation. 
The customised IR contains the backend-specific information, such as the tunable design parameters, and resource and performance models, to describe the characteristics of hardware building blocks. Therefore, the parser of SAMO is responsible for abstracting and unifying the customised IR into the generalised HD-graph, both at node and network levels.

\begin{table}[h]
\centering
\begin{tabular}{|l|ccc|}
\hline
\multicolumn{1}{|c|}{\multirow{2}{*}{\textbf{Variable}}} & \multicolumn{3}{c|}{\textbf{Backend}}                                                                           \\ \cline{2-4} 
\multicolumn{1}{|c|}{}                                   & \multicolumn{1}{c|}{\textit{fpgaConvNet}} & \multicolumn{1}{c|}{\textit{FINN}}         & \textit{HLS4ML}        \\ \hline
\textit{Input Channel Folding}                           & \multicolumn{1}{c|}{Coarse-In}            & \multicolumn{1}{c|}{\multirow{2}{*}{SIMD}} & \multirow{3}{*}{Reuse-Factor} \\ \cline{1-2}
\textit{Kernel Folding}                                  & \multicolumn{1}{c|}{Fine}                 & \multicolumn{1}{c|}{}                      &                        \\ \cline{1-3}
\textit{Output Channel Folding}                          & \multicolumn{1}{c|}{Coarse-Out}           & \multicolumn{1}{c|}{PE}                    &                        \\ \hline
\end{tabular}
\caption{Relationship between backend-specific design parameters and HD-graph optimisation variables.}
\label{tab:backend_variables}
\end{table}

At the node level, the tunable design parameters in the customised IR are mapped to the respective optimisation variables in HD-Graph. This relationship is given in Table~\ref{tab:backend_variables}.
Apart from fpgaConvNet, the relationship between the design parameters and the optimisation variables is not necessarily one-to-one. For example, HLS4ML's degrees of parallelism are summarised within one tunable design parameter, \textit{reuse-factor}, which is mapped to the product of \textit{input channel folding}, \textit{output channel folding} and \textit{kernel folding}.


At the network level, the backends provide their own assertions on the design parameters in order to validate their designs. As such, these assertions in the customised IR are translated to the optimisation constraints of the HD-Graph. Table \ref{tab:backend_constraints} summarises the optimisation constraints required by different backends.

\begin{table}[h]
    \centering
    \begin{tabular}{|l|c|c|c|}
\hline
\textbf{Constraint} & \textbf{fpgaConvNet} & \textbf{FINN} & \textbf{HLS4ML} \\ \hline
\textit{resource} & \cmark & \cmark & \cmark \\ \hline
\textit{channel factor} & \cmark & \cmark & \cmark \\ \hline
\textit{intra channel matching} & \cmark & \cmark & \xmark \\ \hline
\textit{inter channel matching} & \xmark & \cmark & \cmark \\ \hline
    \end{tabular}
    \caption{Constraints over the backends.}
    \label{tab:backend_constraints}
\end{table}

Finally, the parser exposes the resource and latency models to the HD-Graph, if these models have been provided by the backend. For backends which do not contain these models, such as HLS4ML, preliminary latency and DSP predictions are provided, which were obtained analytically.

\subsection{Optimiser: Brute-Force}
SAMO provides three versions of optimisers to exploit the trade-off between the optimisation execution time and the performance of the generated design. Among these optimisers, the Brute-Force search optimiser enumerates all possible values of optimisation variables. During this enumeration, any design point that violates the constraints of the HD-Graph is discarded. The rest of design points are evaluated on the objective function and the optimal one is then identified. The advantage of the Brute-Force optimiser is that it guarantees the identification of the optimal design point, but this is at a cost of lengthy optimisation time.

\subsection{Optimiser: Simulated Annealing}

\begin{algorithm}[h]
\caption{Simulated Annealing Optimisation Algorithm}
\begin{algorithmic}[1]
\State $K$ = $K_{start}$
\Comment{starting temperature}
\State $V = V_{init}$ 
\Comment{initialise as resource minimal}
\While {$K \; < \; K_{min}$}
    \State $V_{prev}$ = $V$
    \Comment{store previous design}
    \State $V$ = random transformation on $V$
    \If{constraints satisfied} 
        \If{ $\psi (V, V_{prev}, K) < x \sim U(0,1)$}
            \State $V$ = $V_{prev}$ 
            \Comment{reject new design}
        \EndIf
    \EndIf
    \State $K = \lambda \cdot K$
    \Comment{reduce temperature}
\EndWhile
\end{algorithmic}
\label{alg:annealing}
\end{algorithm}

Simulated Annealing\cite{reeves1993modern} is a well-known stochastic optimisation algorithm. Its implementation for SAMO is outlined in Algorithm~\ref{alg:annealing}. The algorithm starts by initialising the optimisation variables, $V = \{\bm{C}, \bm{s}^I, \bm{s}^O, \bm{k}\}$ to a resource-minimal state ($V_{init}$), where the computation inside each node is in sequential order and the HD-graph is split completely.

It enters the main optimisation loop where it stores the previous design point and proceeds to perform a random change to the optimisation variables. 
It then evaluates the decision function in Eq.~\eqref{eq:annealing_decision}. If the output is below the decision threshold $x$, which is a random variable sampled from the uniform distribution, or any constraint has been broken, then the current design is discarded and the previous design is kept. This stochastic optimisation process allows the optimiser to navigate out of local minima.
\begin{equation}
    \label{eq:annealing_decision}
    \psi (V, V_{prev},K) = exp\left(\text{min}\left(0, \frac{\mathcal{O}(V_{prev})-\mathcal{O}(V)}{K}\right)\right)
\end{equation}

The algorithm requires two hyper-parameters: $K$ and $\lambda$. $K$ is \textit{temperature}, which starts from $K_{start}$ and decays linearly over iterations until reaching $K_{min}$. The cooling rate, $\lambda$ dictates how fast the temperature decays.

\subsection{Optimiser: Rule-Based}
The Rule-Based optimiser has the same starting point as the Simulated Annealing optimiser, where the optimisation variables are initialised to the resource-minimal state. However, the Rule-Based optimiser solves the optimisation problem using a deterministic method instead, as outlined in Algorithm~\ref{alg:rule}.

\begin{algorithm}[h]
\caption{Ruled-based Optimisation Algorithm}
\begin{algorithmic}[1]
\Procedure{Optimise Partition}{$P$}
\Repeat
    \State $j = \argmax \, t(n_j | s_j^I, s_j^O, k_j), n_j \in P $ 
    \Comment{slowest $j$}
    \State $\Delta = \{ \delta_s^I, \delta_s^O, \delta_k\} \; , \; \Delta > 0$ 
    \Comment{folding increment}
    \State $r_{new} = r(n_j| s_j^I+\delta_s^I, s_j^O+\delta_s^O, k_j+\delta_k) $
    \State $r_{prev} = r(n_j| s_j^I, s_j^O, k_j)$
    \Comment{predict resource}
    \State $ \operatorname*{min}\limits_{\Delta} r_{new} - r_{prev}$ 
    \Comment{smallest resource change}
    \State $s_j^I = s_j^I+\delta_s^I, \;
            s_j^O = s_j^O+\delta_s^O, \;
            k_j = k_j+\delta_k$
\Until{no more resources \textbf{or} fully parallel}
\EndProcedure
\State $V = V_{init}$
\Comment{initialise as resource minimal}
\For{$P_i$ \textbf{in} $H$}
\Comment{optimise partitions independently}
    \State \Call{Optimise Partition}{$P_i$}
\EndFor
\Repeat
\Comment{merge pairs of partitions}
    \For{$P_i$ \textbf{in} $H$}
        \If {$P_i$ meets heuristics}
           \State Merge $P_i$ with $P_{i-1}$ or $P_{i+1}$
        \EndIf
    \EndFor
\Until{no more change in $H$}
\end{algorithmic}
\label{alg:rule}
\end{algorithm}


Firstly, the Rule-based optimiser deals with each partition independently. For each partition, the optimiser identifies the slowest node and increases the optimisation variable which causes the smallest change in terms of the resources. 
This change is denoted as $\Delta$.
Like Simulated Annealing, the change in optimisation variable is propagated throughout the whole HD-graph to fix \textit{intra folding matching} and \textit{inter folding matching}. The above step repeats until the latency of the slowest node cannot be reduced further, either due to reaching resource constraints, or that the slowest node is fully unrolled.

The optimiser then incrementally merges pairs of partitions. This is done by applying heuristics which identify partition merges that are more likely to create an optimal design point. 
These heuristics are based on the following properties of the partition, 
\begin{itemize}
    \item is memory-bound
    \item the slowest node is fully unrolled
    \item latency is smaller than reconfiguration time
\end{itemize}
Once all the identified partitions cannot be merged further, the optimiser terminates. 

\subsection{Exporter: customised IR to HD-graph}
The optimised HD-graph is transformed back to the customised IR belonging to the respective backend, and the design parameters of the hardware building blocks are configured with the values of the corresponding optimisation variables. This optimised IR can then be used to synthesise and generate bitstreams for the target platform.

\section{Evaluation}


The proposed SAMO framework is evaluated in terms of its ability to identify high-performance designs under both throughput and latency objectives, as well as with respect to its scalability.


\begin{table}[h]
\setlength{\tabcolsep}{4pt}
\centering
\begin{tabular}{@{}ccccc@{}}
\toprule
\textbf{Task} & \textbf{Network} & \textbf{No. Conv} & \textbf{No. Dense} & \textbf{Params}  \\ \midrule
Jet Tagging & \textit{3-layer}  & 0 & 4 & 4K\\ \midrule
Hand Gestures & \textit{MPCNN}  & 3 & 2 & 70K\\ \midrule
\multirow{2}*{MNIST} & \textit{TFC}  & 0 & 4 & 59K \\
~ & \textit{LeNet} & 2 & 2 & 430K\\ \midrule
CIFAR-10 & \textit{CNV} & 6 & 3 & 1.543M \\ \midrule
\multirow{3}*{ImageNet} & \textit{VGG11} & 8 & 3 & 132.854M\\ 
~ & \textit{MobileNetV1} & 27 & 1  & 4.209M\\ 
~ & \textit{ResNet50}  & 52 & 1 & 25.371M\\ \bottomrule
\end{tabular}
\caption{Model zoo for evaluation.}
\label{tab:cnn_models}
\end{table}

The CNN models used in the evaluation are sourced from the design examples provided by the backends. These models are outlined in Table~\ref{tab:cnn_models}. In terms of target platforms, ZedBoard, ZC706 and U250 are selected in order to evaluate the framework over resource profiles that span from embedded systems to high-end server-graded FPGA device. 


\begin{figure*}[t]
    \centering
    \begin{subfigure}{0.4\textwidth}
        \includegraphics[width=\textwidth]{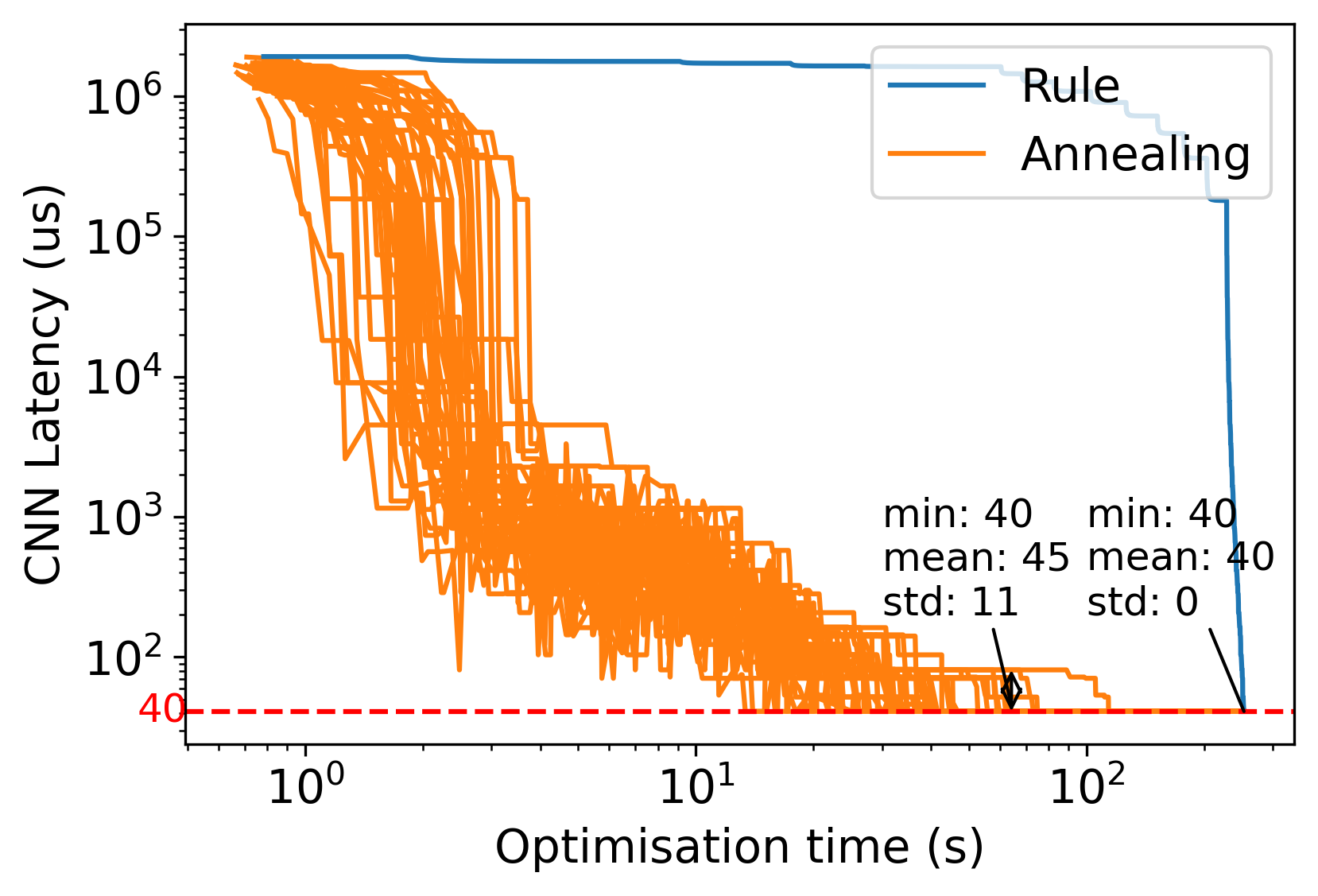}
        \caption{CNV (w1a1)}
    \end{subfigure}
    \begin{subfigure}{0.4\textwidth}
        \includegraphics[width=\textwidth]{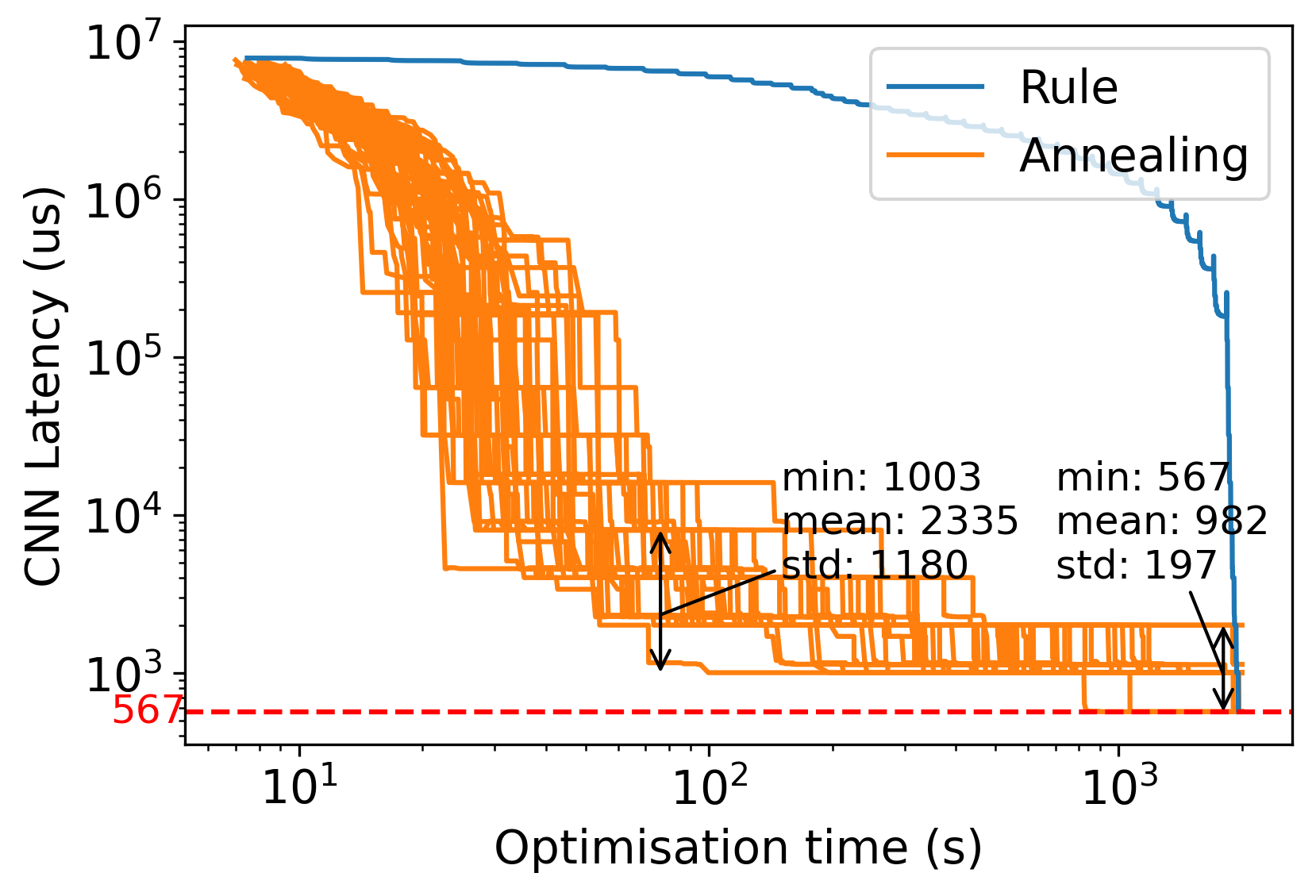}
        \caption{MobileNetV1 (w4a4)}
    \end{subfigure}
    \caption{Comparison between the Simulated Annealing and Rule-Based optimisers. Networks are mapped to a U250 using the FINN backend with a latency objective. For the Rule-Based optimiser, the horizontal red dashed line represents the CNN latency of the final design point. The Simulating Annealing optimiser restarts for 50 times with different random seeds and the minimum, mean and standard deviation of these 50 runs are annotated.}
    \label{fig:optimiser_comparison}
\end{figure*}

\subsection{Scalability}
As the Brute-Force optimiser exhaustively explores the design space, it is used to explain the scale of the optimisation problem. The results are shown in Table \ref{tab:brute_force}, which highlights the size of the design space, the number of design points explored per second, and the estimated time to evaluate the whole design space.

\begin{table}[h]
\setlength{\tabcolsep}{4pt}
\centering
\begin{tabular}{@{}ccccc@{}}
\toprule
\textbf{Network} & \textbf{Backend}& \textbf{Size}& \textbf{Points /s}& \textbf{Est. Time} \\ \midrule
\multirow{3}{*}{\textit{3-layer}} & HLS4ML & 1.02x$10^9$    & 1334 & 9 days     \\
 & FINN                                    & 5.08x$10^{8}$  & 120  & 5 days     \\
 & fpgaConvNet                             & 2.58x$10^{12}$ & 378  & 22 decades \\ \midrule
\multirow{3}{*}{\textit{TFC}} & HLS4ML & 4.99x$10^9$    & 1686 & 34 days \\ 
 & FINN                                & 1.02x$10^{11}$ & 120  & 3 years \\ 
 & fpgaConvNet                         & 5.32x$10^{13}$ & 377  & 45 centuries \\ \midrule
\multirow{3}{*}{\textit{LeNet}} & HLS4ML & 4.85x$10^{11}$ & 1443 & 11 years \\ 
 & FINN                                  & 8.02x$10^{13}$ & 85   & 299 centuries \\ 
 & fpgaConvNet                           & 8.02x$10^{13}$ & 285  & 89 centuries\\ \midrule
\multirow{3}{*}{\textit{CNV}} & HLS4ML & 8.05x$10^{30}$ & 703 & 4x$10^{18}$ centuries \\
 & FINN                                & 9.98x$10^{37}$ & 40  & 8x$10^{26}$ centuries \\
 & fpgaConvNet                         & 2.00x$10^{42}$ & 125 & 5x$10^{30}$ centuries \\ \bottomrule
\end{tabular}
\caption{The design space and the estimated exploration time using the Brute-Force optimiser.}
\label{tab:brute_force}
\end{table}

The table demonstrates the large design space defined by the Streaming Architecture accelerators. Apart from the \textit{3-layer} network, exploring the whole design space is intractable, with some tests being estimated to take centuries to complete. Upon further inspection, it is evident that the size of the design space mainly scales with the number of layers inside a network and the number of channels inside each layer.


The size of design space also varies across different backends.
When all available backends target the same model, fpgaConvNet typically has the largest design space because it supports three design parameters (\textit{Coarse-In}, \textit{Coarse-Out} and \textit{Fine}) in each hardware building block, while other backends support one or two only. In addition, the rate of evaluating these design points is also a limitation due to the complex latency and resource model evaluation as well as constraint evaluation. 

\subsection{Choice of Optimisers}

\begin{table*}[t]
\centering
\begin{tabular}{@{}cccccccccccccc@{}}
\toprule
\multirow{2}{*}{\textbf{Backend}} & \multirow{2}{*}{\textbf{\begin{tabular}[c]{@{}c@{}}Network/\\ Precision\end{tabular}}} & \multicolumn{3}{c}{\textbf{Partitions}} & \multicolumn{3}{c}{\textbf{Latency (ms/batch)}} & \multicolumn{3}{c}{\textbf{Throughput (img/s)}} & \multicolumn{3}{c}{\textbf{Resource (\%)}} \\ \cmidrule(l){3-14} 
 &  & \textit{init.} & \textit{lat.} & \textit{thr.} & \textit{init.} & \textit{lat.} & \textit{thr.} & \textit{init.} & \textit{lat.} & \textit{thr.} & \textit{init.} & \textit{lat.} & \textit{thr.} \\ \midrule
\multirow{2}{*}{fpgaConvNet} & LeNet/w16a16 & 1 & 1 & 3 & 16.0 & 2.0 & 283.9 & 62.5 & 500.0 & 901.6 & 50.8 & 50.5 & 83.5 \\
 & CNV/w16a16 & 1 & 6 & 8 & 289.0 & 421.7 & 1304.7 & 3.5 & 2.4 & 196.2 & \textcolor{red}{120.5} & 64.5 & 82.0 \\ \midrule
\multirow{3}{*}{FINN} & MPCNN/w4a4 & 1 & 2 & 3 & 10.0 & 84.5 & 432.1 & 100 & 11.8 & 592.4 & \textcolor{red}{48.1} & 44.4 & 47.4 \\
 & CNV/w1a1 & 1 & 1 & 1 & 289.0 & 0.3 & 73.7 & 3.5 & 3472.2 & 3472.2 & 18.7 & 33.0 & 33.0 \\
 & MobileNetV1/w4a4 & 1 & 5 & 13 & 513.8 & 475.3 & 11347.0 & 1.9 & 2.1 & 22.6 & \textcolor{red}{187.1} & 63.8 & 82.8 \\ \bottomrule
\end{tabular}
\caption{Comparison of optimisation results for the Rule-Based optimiser against unoptimised designs (\textit{init.}) targeting both throughput (\textit{thr.}) and latency (\textit{lat.}) objectives for a ZedBoard device. The batch size for the throughput objective is 256. Resource is the average utilisation across DSP, BRAM and LUT. Designs with resources in \textcolor{red}{red} violate constraints.}
\label{tab:objective_with_partitions}
\end{table*}

Beyond the Brute-Force optimiser, SAMO also provides the Simulated Annealing optimiser and Rule-Based optimiser to traverse the design space and deal with it's large cardinality. The two have very different properties; the Rule-Based optimiser is a deterministic algorithm, and always produces the same design for a network-platform pair. On the other hand, due to the stochasticity of the annealing algorithm, results from the Simulated Annealing optimiser may vary from run to run.

Fig.~\ref{fig:optimiser_comparison} demonstrates the performance of these two optimisers by reporting the optimisation time and achieved performance. 
For the Simulated Annealing runs, the hyper-parameter temperature $K$ is initialised as 1000 and reduced by 2\% every iteration until reaching the minimum temperature $K_{min}$=1. It is then left to run at minimum temperature and for the same time budget as the Rule-Based optimiser.

The distribution across runs for Simulated Annealing is calculated and compared with the Rule-Based results. 
For CNV, all 50 runs of the Simulated Annealing optimiser give the exactly the same latency as the Rule-Based optimiser and converge quicker, suggesting that Simulated Annealing is able to find optimal designs much more rapidly.

However, this is not the case when it comes to the optimisation of a wider and deeper\footnote{``wide" and ``deep" are used here to describe the number of channels and the number of layers.} network, \textit{MobileNetV1}, where the Simulated Annealing runs did not converge. Considering each run of the optimiser takes about 33 minutes to complete, the stochasticity of the annealing algorithm becomes a major drawback for repeatable high performance designs.

To summarise, the results suggest that the Brute-Force, Simulated Annealing and Rule-Based Optimiser should be used for small, medium, and large networks respectively. The Brute-Force optimiser guarantees identification of an optimal design point, but its lengthy search time makes it only feasible for really small networks. Both the Simulated Annealing optimiser and the Rule-Based optimiser efficiently explore the design space at the risk of being stuck at a local minimum. However, the Rule-Based optimiser performs better whilst handling larger networks,  as the randomness of Simulated Annealing makes it sub-optimal when the same time budget is considered.


\subsection{Discussion on Partitioning}

As described in Section~\ref{subsec:Objective}, SAMO supports both latency and throughput objectives for optimisation. By introducing partitioning into the design space, these two objectives can lead to very different hardware outcomes. This section evaluates the difference in throughput and latency driven designs for both high and low-end devices.

\begin{figure}[h]
    \centering
    \includegraphics[width=0.4\textwidth]{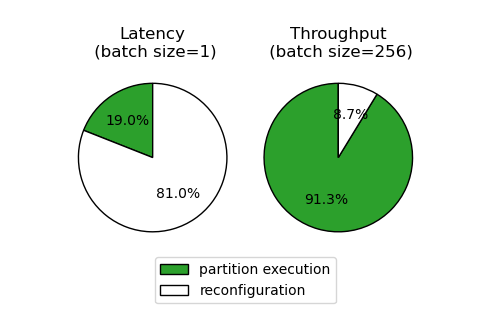}
    \caption{Comparison of percentage of time spent in partition execution and reconfiguration for a latency and throughput objective for MobileNetV1 from Table~\ref{tab:objective_with_partitions}. }
    \label{fig:partition_pie}
\end{figure}

Table~\ref{tab:objective_with_partitions} demonstrates the capability of SAMO to identify the design points for a resource-constrained device. Both throughput and latency objectives are compared to an unoptimised design whose optimisation variables are all set to 1. 

The first benefit of introducing partitioning is the ability to overcome resource constraints. For \textit{CNV} for fpgaConvNet, and \textit{MPCNN} and \textit{MobileNetV1} for FINN, the unoptimised designs exceed the available resources for the ZedBoard. By splitting the HD-Graph into multiple sub-graphs, SAMO is able to overcome this constraint for the latency and throughput optimised designs. This is at the cost of increased latency needed for reconfiguration, as illustrated in Fig.~\ref{fig:partition_pie}. And in all cases where the unoptimised design fits, SAMO is able to find an improved design.

Partitioning also enables much greater throughput to be achieved than being constrained to a single partition. In Table~\ref{tab:objective_with_partitions}, it can be seen that for \textit{LeNet} for fpgaConvNet, the throughput for the throughput-driven design is nearly double that of the latency-driven one. This is in part due to the ability to amortise the cost of reconfiguration through large batch sizes, which is also illustrated in Fig.~\ref{fig:partition_pie}. Here the throughput-driven design spends significantly more time executing hardware than reconfiguring. It is also observed that throughput-driven designs lead to more efficient use of the hardware per partition, with much higher resource utilisation compared to latency-driven designs.

\begin{figure}[h]
    \centering
    \includegraphics[width=0.49\textwidth]{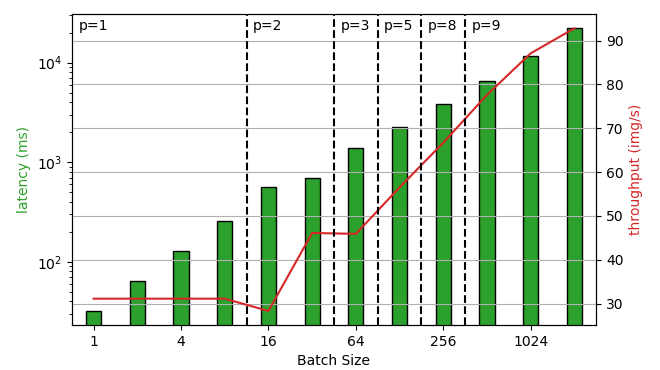}
    \caption{Comparison of Throughput and Latency using different batch sizes for a VGG11 network targetting a U250 device using the fpgaConvNet backend. \textit{p} indicates the number of partitions.}
    \label{fig:batch_size}
\end{figure}





Exploring throughput-driven designs further, different batch sizes are used in Fig.~\ref{fig:batch_size} to show achievable throughput for a U250 device deploying \textit{VGG11} using the fpgaConvNet backend. This figure highlights that partitioning is not only a mechanism for satisfying resource constraints, but also a way of further improving throughput. As the batch size is increased, more partitions are used to increase throughput.  

\subsection{Comparison with Existing Designs}

Table~\ref{tab:baseline_comparison} compares designs generated by SAMO with example designs provided by authors of each backend, where the design parameters of the hardware building blocks are manually tuned. The results highlight the power of SAMO's automated design space exploration, as it can achieve the same or higher performance compared with a hand-crafted method, demonstrating improvements of 4-20x in performance across different backends.

\begin{table}[h]
    \centering
    \begin{tabular}{@{}ccccc@{}}
\toprule
\toprule
    \multirow{2}*{{\textbf{Backend}}} & \multirow{2}*{{\textbf{Platform}}} & \multirow{2}*{{\textbf{Network/Precision}}} & \multicolumn{2}{c}{\textbf{Latency (us)}} \\
    ~ & ~ & ~ & \textit{baseline} & \textit{SAMO}  \\
    \midrule
    \midrule
    HLS4ML\tablefootnote{https://github.com/fastmachinelearning/hls4ml} & U250 & 3-layer/w16a16 & 0.001 & 0.001 \\
    \midrule
   \multirow{2}*{\makecell{fpgaConvNet\\\cite{venieris_fpgaconvnet_2016}}} & \multirow{2}*{Zedboard} & LeNet/w16a16 & 7917.0 & \textbf{2000.0} \\
    ~ & ~ & MPCNN/w16a16 & 3919.0 & \textbf{180.0} \\
    \midrule
    \multirow{3}*{FINN\tablefootnote{https://github.com/Xilinx/finn-examples}} & \multirow{3}*{U250} & CNV/w1a1 & 163.8 & \textbf{41.0}  \\
    ~ & ~ & MobileNetV1/w4a4 & 567.9 & 567.9\\
    ~ & ~ & ResNet-50/w1a2 & 4515.8 & \textbf{3081.3} \\
    \bottomrule
    \bottomrule
    \end{tabular}
    \caption{Comparison with baseline designs. Latency is predicted by the backend performance models.\\
    }
    \label{tab:baseline_comparison}
\end{table}

Furthermore, the optimised designs are synthesised, and their utilisation is reported in Table~\ref{tab:rsc}. This table highlights the differences between predicted resources from the backends' models, and validates SAMO's ability to generate usable designs. It can be seen that all designs fit within their target platforms.

\begin{table}[h]
\centering
\begin{tabular}{@{}ccccccc@{}}
\toprule
\toprule
\textbf{Backend} & \multicolumn{2}{c}{HLS4ML} & \multicolumn{2}{c}{fpgaConvNet} & \multicolumn{2}{c}{FINN}\\ \midrule
\textbf{Network/Platform} & \multicolumn{2}{c}{3-layer/U250} & \multicolumn{2}{c}{LeNet/Zedboard} & \multicolumn{2}{c}{CNV/U250} \\ \midrule
\midrule
\textbf{Resource (\%)} & \textit{Pred.}  & \textit{Synth.} & \textit{Pred.}  & \textit{Synth.} & \textit{Pred.} & \textit{Synth.}\\ \midrule
\textbf{DSP}  & 17.3 & 17.3 &   4.1 & 13.6 & 0.0 & 0.0  \\
\textbf{BRAM} &  0.0 & 17.6 & 177.9* & 91.1 & 0.0 & 0.0  \\
\textbf{LUT}  &  0.0 &  1.1 &  94.6 & 46.7 & 8.1 & 13.3 \\ 
\textbf{FF}   &  0.0 &  3.5 &  15.9 &  7.3 & 0.0 & 0.0  \\ 
\bottomrule
\bottomrule
\end{tabular}
\caption{Resources comparison of model prediction and post-synthesis results, of optimised designs identified by SAMO.\\
\textit{*BRAM constraint was relaxed due to over-estimation.}}
\label{tab:rsc}
\end{table}

\section{Conclusion \& Future Work}

This paper presents the SAMO framework, an open-source Streaming Architecture Mapping Optimiser, which serves as a powerful tool for CNN Accelerator designers.
The framework has been integrated with popular open-source Streaming Architectures in order to prove it's ability in achieving high performance designs across a range of CNN networks and FPGA platforms. 
The potential of the proposed optimisers (Simulated Annealing and Rule-Based) are demonstrated, with considerable gains in performance observed.
This framework can be seen as a launchpad for further research into CNN-FPGA co-design, with potential for use in the exploration of Neural Architecture Search (NAS), as well as exploring improved optimisation methods.


\bibliographystyle{IEEEtran}
\typeout{}
\bibliography{bibliography}

\begin{thebibliography}{10}
\providecommand{\url}[1]{#1}
\csname url@samestyle\endcsname
\providecommand{\newblock}{\relax}
\providecommand{\bibinfo}[2]{#2}
\providecommand{\BIBentrySTDinterwordspacing}{\spaceskip=0pt\relax}
\providecommand{\BIBentryALTinterwordstretchfactor}{4}
\providecommand{\BIBentryALTinterwordspacing}{\spaceskip=\fontdimen2\font plus
\BIBentryALTinterwordstretchfactor\fontdimen3\font minus
  \fontdimen4\font\relax}
\providecommand{\BIBforeignlanguage}[2]{{%
\expandafter\ifx\csname l@#1\endcsname\relax
\typeout{** WARNING: IEEEtran.bst: No hyphenation pattern has been}%
\typeout{** loaded for the language `#1'. Using the pattern for}%
\typeout{** the default language instead.}%
\else
\language=\csname l@#1\endcsname
\fi
#2}}
\providecommand{\BIBdecl}{\relax}
\BIBdecl

\bibitem{venieris_fpgaconvnet_2018}
S.~I. Venieris and C.-S. Bouganis, ``{fpgaConvNet}: Mapping regular and
  irregular convolutional neural networks on fpgas,'' \emph{IEEE Transactions
  on Neural Networks and Learning Systems}, vol.~30, no.~2, 2019.

\bibitem{duarte_fast_2018}
J.~Duarte, S.~Han, P.~Harris, S.~Jindariani, E.~Kreinar, B.~Kreis, J.~Ngadiuba,
  M.~Pierini, R.~Rivera, N.~Tran, and Z.~Wu, ``Fast inference of deep neural
  networks in {FPGAs} for particle physics,'' \emph{Journal of
  Instrumentation}, vol.~13, no.~07, 2018.

\bibitem{blott2018finn}
M.~Blott, T.~B. Preu{\ss}er, N.~J. Fraser, G.~Gambardella, K.~O’brien,
  Y.~Umuroglu, M.~Leeser, and K.~Vissers, ``{FINN-R}: An end-to-end
  deep-learning framework for fast exploration of quantized neural networks,''
  \emph{ACM Transactions on Reconfigurable Technology and Systems}, vol.~11,
  no.~3, 2018.

\bibitem{venieris2018toolflows}
S.~I. Venieris, A.~Kouris, and C.-S. Bouganis, ``Toolflows for mapping
  convolutional neural networks on fpgas: A survey and future directions,''
  \emph{ACM Computer Surveys}, vol.~51, no.~3, 2018.

\bibitem{jouppi2017datacenter}
N.~P. Jouppi, C.~Young, N.~Patil, D.~Patterson, G.~Agrawal, R.~Bajwa, S.~Bates,
  S.~Bhatia, N.~Boden, A.~Borchers \emph{et~al.}, ``In-datacenter performance
  analysis of a tensor processing unit,'' in \emph{Proceedings of the 44th
  annual international symposium on computer architecture}, 2017.

\bibitem{vink2020caffe}
D.~A. Vink, A.~Rajagopal, S.~I. Venieris, and C.-S. Bouganis, ``Caffe barista:
  Brewing caffe with fpgas in the training loop,'' in \emph{International
  Conference on Field-Programmable Logic and Applications}, 2020.

\bibitem{umuroglu2017finn}
Y.~Umuroglu, N.~J. Fraser, G.~Gambardella, M.~Blott, P.~Leong, M.~Jahre, and
  K.~Vissers, ``{FINN}: A framework for fast, scalable binarized neural network
  inference,'' in \emph{International Symposium on Field-Programmable Gate
  Arrays}, 2017.

\bibitem{wei2017automated}
X.~Wei, C.~H. Yu, P.~Zhang, Y.~Chen, Y.~Wang, H.~Hu, Y.~Liang, and J.~Cong,
  ``Automated systolic array architecture synthesis for high throughput cnn
  inference on fpgas,'' in \emph{Proceedings of the 54th Annual Design
  Automation Conference 2017}, 2017.

\bibitem{zhang_optimizing_2015}
C.~Zhang, P.~Li, G.~Sun, Y.~Guan, B.~Xiao, and J.~Cong, ``Optimizing
  {FPGA}-based accelerator design for deep convolutional neural networks,'' in
  \emph{International Symposium on Field-Programmable Gate Arrays}, 2015.

\bibitem{dicecco2016caffeinated}
R.~DiCecco, G.~Lacey, J.~Vasiljevic, P.~Chow, G.~Taylor, and S.~Areibi,
  ``Caffeinated fpgas: Fpga framework for convolutional neural networks,'' in
  \emph{International Conference on Field-Programmable Technology}, 2016.

\bibitem{li2016high}
H.~Li, X.~Fan, L.~Jiao, W.~Cao, X.~Zhou, and L.~Wang, ``A high performance
  fpga-based accelerator for large-scale convolutional neural networks,'' in
  \emph{International Conference on Field Programmable Logic and Applications},
  2016.

\bibitem{zhang2018dnnbuilder}
X.~Zhang, J.~Wang, C.~Zhu, Y.~Lin, J.~Xiong, W.-m. Hwu, and D.~Chen,
  ``Dnnbuilder: an automated tool for building high-performance dnn hardware
  accelerators for fpgas,'' in \emph{International Conference on Computer-Aided
  Design}, 2018.

\bibitem{venieris_fpgaconvnet_2016}
S.~I. Venieris and C.-S. Bouganis, ``{fpgaConvNet}: A framework for mapping
  convolutional neural networks on fpgas,'' in \emph{2016 IEEE 24th Annual
  International Symposium on Field-Programmable Custom Computing Machines
  (FCCM)}, 2016.

\bibitem{faraone2018customizing}
J.~Faraone, G.~Gambardella, D.~Boland, N.~Fraser, M.~Blott, and P.~H. Leong,
  ``Customizing low-precision deep neural networks for fpgas,'' in
  \emph{International Conference on Field Programmable Logic and Applications},
  2018.

\bibitem{alonso2021elastic}
T.~Alonso, L.~Petrica, M.~Ruiz, J.~Petri-Koenig, Y.~Umuroglu, I.~Stamelos,
  E.~Koromilas, M.~Blott, and K.~Vissers, ``{Elastic-DF}: Scaling performance
  of {DNN} inference in {FPGA} clouds through automatic partitioning,''
  \emph{ACM Transactions on Reconfigurable Technology and Systems}, vol.~15,
  no.~2, 2021.

\bibitem{aarrestad2021fast}
T.~Aarrestad, V.~Loncar, N.~Ghielmetti, M.~Pierini, S.~Summers, J.~Ngadiuba,
  C.~Petersson, H.~Linander, Y.~Iiyama, G.~Di~Guglielmo \emph{et~al.}, ``Fast
  convolutional neural networks on fpgas with hls4ml,'' \emph{arXiv preprint},
  2021.

\bibitem{reeves1993modern}
C.~R. Reeves, Ed., \emph{Modern Heuristic Techniques for Combinatorial
  Problems}.\hskip 1em plus 0.5em minus 0.4em\relax USA: John Wiley \& Sons,
  Inc., 1993.

\end{thebibliography}

\end{document}